\documentclass[12pt,aps,bm,epsfig,amssymb]{article}
\usepackage{epsfig,amsmath,amsfonts,float}
\newcommand{\R}[0]{{\mathbb{R}}}
\newcommand{\Z}[0]{{\mathbb{Z}}}
\newcommand{\cH}{{\cal H}}
\newcommand{\cN}{{\cal N}}

\newcommand{\cC}{{\cal C}}
\newcommand{\cK}{{\cal K}}
\def\C{{\mathbb{C}}}
\def\Z{{\mathbb{Z}}}
\def\R{{\mathbb{R}}}

\newtheorem{Lemma}{Lemma}

\newtheorem{Theorem}{Theorem}

\newcommand{\bef}{\begin{figure}}
\newcommand{\eef}{\end{figure}}

\newcommand{\bei}{\begin{itemize}}
\newcommand{\eei}{\end{itemize}}
\newcommand{\bea}{\begin{eqnarray}}

\newcommand{\eea}{\end{eqnarray}}
\newcommand{\bequ}{\begin{equation}}
\newcommand{\eequ}{\end{equation}}

\DeclareRobustCommand\openone{\leavevmode\hbox{\small1\normalsize\kern-.33Spin-1/2 particles moving on a 2D lattice with 
 nearest-neighbor interactions can realize an autonomous quantum computerem1}}

\begin{document}

\title{\Large \textbf{Spin-1/2 particles moving on a 2D lattice with 
 nearest-neighbor interactions can realize an autonomous quantum computer}}

\author{Dominik Janzing\\
{\small Institut f\"{u}r Algorithmen und Kognitive Systeme,} \\{\small 
Universit\"{a}t Karlsruhe, Am Fasanengarten 5, 76 131 Karlsruhe, Germany
}
\\
{\small and Institut f\"{u}r Quantenoptik und Quanteninformation}\\ {\small
der \"{O}sterreichischen Akademie der Wissenschaften, 
6020 Innsbruck, Austria } 
}

\date{June 30, 2005}

\maketitle

%
\begin{abstract}
What is the simplest Hamiltonian which can implement quantum computation
without requiring any control operations during the computation process?
In a previous paper we have constructed a 10-local finite-range interaction
among qubits on a 2D lattice having this property. 
Here we show that pair-interactions among
qutrits on a 2D lattice are sufficient, too, and can also implement 
an {\it ergodic computer} where the result can be read out 
from the time average state after  some post-selection 
with high success probability. 

Two of the 3 qutrit states 
are given by the two levels of a spin-1/2 particle located at a 
specific lattice site, the third state is its absence.
Usual hopping terms together with an attractive force 
among adjacent particles induce a coupled quantum walk where
the particle spins are subjected to spatially inhomogeneous
interactions implementing holonomic quantum computing. 
The holonomic method ensures that the implemented circuit 
does not depend on the time needed for the walk.

Even though the implementation of the required 
type of spin-spin interactions
is currently unclear, the model shows that quite simple
Hamiltonians  are powerful enough to allow for universal quantum computing in 
a closed physical 
system. 
\end{abstract}

\section{Introduction}

To understand which sets of quantum control operations are sufficient for quantum computing
is still an important issue of research. 
Whereas the standard model of the quantum computer
is based on one- and two-qubit unitaries there is meanwhile a large number of alternative proposals, e.g. computing by measurements only
 \cite{RB00,CompMess} 
or adiabatic computing \cite{Kaminsky,FarhiAdiabatic, 
WilliamsAdiabatic,AdiabaticQC,Paweladiabatic}.
The latter model encodes a  computational problem
into an interaction such that the ground state of the Hamiltonian
indicates the solution. In order to understand 
the power of adiabatic computing it is particularly interesting to  know to 
what extent
the  transition into the ground state of simple interactions 
can already be a non-trivial computation process. 
Since it has been shown that every problem in the 
complexity class QMA can be encoded in a nearest-neighbor interaction
of qubits located on a 2D lattice \cite{Oliveira} it is clear that, indeed,
relatively simple Hamiltonians are sufficient for adiabatic
computing. This highlights the computational power of 
simple Hamiltonians from one point of view. 
Another aspect of the
``computational power of Hamiltonians'' has been studied
in models where the natural time evolution is interspersed
by external control operations (see e.g. 
\cite{Martineffizient,Timesteps,PawelDiss,Bremner}). 
The intention of the present paper is to understand 
to what extent models with simple interactions can already be
autonomous devices with full quantum computing power 
in the sense that quantum computation reduces to the following
protocol:
(1) prepare an initial state in the computational basis
which contains the data,
(2) wait for a sufficiently long time, and (3) measure a sufficiently large
set of qubits in the computational basis. 
In \cite{Ergodic} we have constructed a Hamiltonian satisfying 
these conditions with the additional feature that the readout
need not necessarily be performed within a specific time interval;
in our ``ergodic quantum computer'' the result is also present
in the time average of the dynamics after some simple
post-selection with high success probability. 
 Even though this property 
may be of minor practical relevance, we considered it as a necessary
feature of an {\it autonomous} computer since one  would otherwise require 
a clock as an additional device. Here we construct an interaction that
is even simpler than the one constructed in \cite{Ergodic}
since the latter requires 10-qubit interactions.

The motivation to consider autonomous quantum computers  is twofold.
First, it could  trigger new ideas how  to reduce the 
set of necessary control operations in current implementation proposals
by using the ``natural power'' of the interactions. In addressing this issue,
the ergodic model 
defines
only an extreme case;  realistic perspectives 
for quantum computing could arise by combining
it with more conventional
approaches. 
Second it is an interesting fundamental issue in the
thermodynamics of computation how to realize computation in closed 
physical systems. Benioff, Feynman, and Margolus have already presented
such Hamiltonian computers \cite{Benioff,Feynman:85,Margolus:90}.  
Margolus' asynchronous Hamiltonian cellular automaton (CA) 
has the appealing feature 
to be 
lattice-symmetric. In other words, it is driven by a
 finite-range interaction among qubits\footnote{In \cite{Biafore}
it is furthermore argued that one of this
clock Hamiltonians for CAs in one dimension is quite close to real 
solid state interactions.}
located on a 2D lattice such that the total Hamiltonian
is invariant under discrete translations.
The synchronization of the CA is realized by a kind of spin wave propagating
along the lattice and triggering the update of the cells
according to some computationally universal update rules which are not
specified any further in \cite{Margolus:90}.  
However, its clock wave has to start in an uncertain {\it position}
in order to obtain a well localized {\it momentum} distribution with
mainly positive momenta. A localized wave front would also propagate 
backwards and would therefore not trigger a correct computation.
In \cite{Ergodic}  we have chosen the same synchronization scheme  
but
 we start with a localized wave front since we allow only
for preparations of {\it basis} states. The fact that the
dispersion of the wave front leads to completely 
undefined  computation steps 
is irrelevant since the {\it time average} of 
the  dynamics encodes the correct result.
The feature of our ergodic model to show the correct computation result
also in the time average was hence only a nice byproduct 
of the fact that we must use a concept which works 
with a strange kind of clock:
The latter starts with a well-defined time but then it
counts 
backwards and forward with completely undefined counting speed. 
Note that the time average
can also be considered as the ``generalized final state''
of our computer since a final output state in the usual sense cannot exist
for finite dimensional Hamiltonian models.  
 

To ask for a simple finite range Hamiltonian that is universal 
for quantum computation is in some sense similar to asking 
for a simple computationally universal quantum cellular automaton 
(for some proposals see \cite{RaussendorfCA,Shepherd}) 
with the only difference of
considering update rules which change the state
only in an ``infinitesimal'' way. 
However, local interactions in lattices have typically the
property to spread the information into increasingly large regions, whereas
it is possible  to construct update rules for cellular automata
that work by propagating the information forward column by column 
\cite{RaussendorfCA}. This apparent difference between discrete and 
continuous time
dynamics can already be explained with
the translation operator on $n$ qubits: the cyclic shift is a unitary that 
can 
be achieved by
local update rules \cite{WernerCA}, 
whereas the Hamiltonian obtained
from the logarithm  of the shift contains interactions between
{\it distant} qubits. Hamiltonians with finite interaction
length will typically spread the information  
over the lattice. 
Therefore,
one has to be more modest and 
demand that the correct result can only be present with high probability. 

The paper is organized as follows. 
In section~\ref{Syn} we explain the hardware of our model 
consisting of a chain of  atoms on a 2D square lattice. 
We construct a nearest-neighbor  interaction which leads 
via a simple effective Hamiltonian
to a
coupled random walk that will later trigger the implementation of gates
acting on the atom spins.    
In section~\ref{Holo} we explain 
how to arrange spin-spin interactions among the atoms such that
they implement logical gates based on holonomic quantum 
computing. 
In section~\ref{Complete} we describe the complete Hamiltonian
of our computing device. 
In section~\ref{Sec:Diag} we show that the time 
evolution of the effective Hamiltonian is solvable since 
it can be transformed into a quasifree evolution of fermions. 
Based on this solution, 
we estimate
the time required for the computation  process. 
In section~\ref{Impl} we briefly sketch some ideas on the realization. 
Section~\ref{Sec:CA} describes how 
to construct the Hamiltonian such that it implements 
a universal quantum cellular automaton. 
This is to obtain programmable hardware. 
In the appendix we present an alternative model for the atom propagation
which is caused by an even simpler Hamiltonian since the
nearest-neighbor interactions in the
 model presented
in the main part involves also {\it diagonal neighbors} in the lattice.
In contrast, the model in the appendix uses only {\it nearest} neighbors in 
a strict sense. 
However, the 
corresponding Hamiltonian dynamics will not be solvable.
We will therefore consider 
an incoherent analogue and show that the corresponding
classical random walk would trigger the implementation of gates
in the desired way. 
For the coherent model, we can only 
 conjecture that an appropriate propagation is achieved.

\section{The synchronization Hamiltonian}

\label{Syn}

\begin{figure}
\centerline{\epsfbox[0 0 222 193]{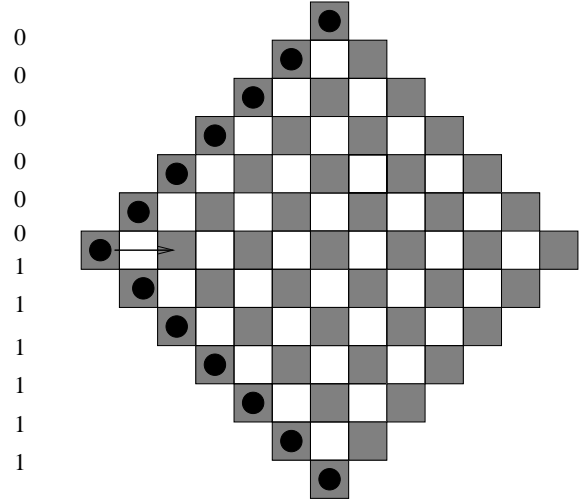}} 
\caption{{\small Initial atom configuration for a $7\times 7$ lattice. 
The digits at the left
indicate the corresponding binary string
which will be introduced in section~\ref{Sec:Diag}
and the arrow shows
the only possible motion.\label{DiagonalPositionen}}}
\end{figure}

Now we 
present a model of a 2D lattice with a coupled propagation of
a chain of atoms which will later be the synchronization mechanism
of the computer. While propagating along the horizontal direction
(see Fig.~\ref{DiagonalPositionen}) the spin of the atoms
(which represent the logical states) will be subjected to  spatially
inhomogeneous 
interactions that implement the desired gates. 
It is important that the chain does not tear since 
the spin-spin interactions to be described later 
will only be active between adjacent atoms. Furthermore,
it is important for the synchronization that the chain
remains connected. This will become clear in section~\ref{Holo}.

\begin{figure}
\centerline{\epsfbox[0 0 222 193]{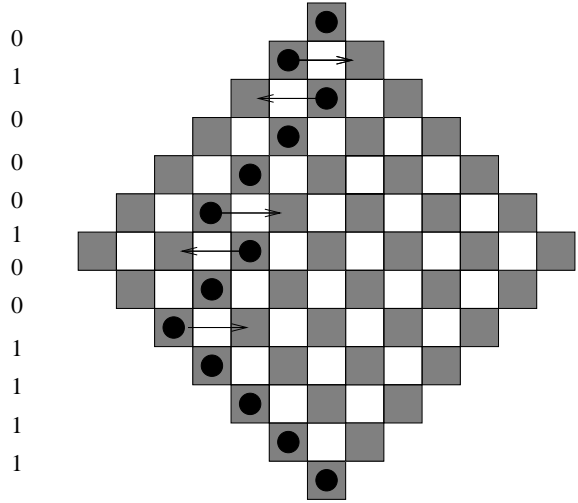}}
\caption{{\small 
Possible atom configuration with its corresponding binary string
(as explained in section~\ref{Sec:Diag}) 
and the possible motions.\label{Possible<}}}
\end{figure}

We will now describe the synchronization Hamiltonian
in detail and show that it generates
a diagonalizable quantum  walk.
The lattice and the initial atom configuration 
are shown in Fig.~\ref{DiagonalPositionen}. The lattice sites are given by
$n\times n$ black fields of the modified chess-board in 
Fig.~\ref{DiagonalPositionen}.
The $2n-1$ atoms are only allowed to move  forward or backward
along the horizontal direction, i.e., along the rows.
Each atom can only hop from a black site to either the next or 
the previous black site in the row.  
Furthermore, the atoms can only move such that the chain does not tear, i.e.,
atoms in adjacent rows are always diagonal neighbors. 
Fig.~\ref{Possible<} shows a possible configuration.

Now we describe the Hamiltonian  
that allows only  collective motions of the atoms respecting these rules.
The grid of black fields is labeled by $(i,j)$ with $i,j=1,\dots,n$. 
The sites $(1,j)$ with $j=1,\dots,n$ indicate the diagonal line leading from
the leftmost field to the uppermost one and sites $(i,1)$ with $i=1,\dots,n$ 
indicate
the diagonal line from the leftmost field to the lower-most one. 
We consider the Hilbert space 
of all atom configurations as a subspace of 
an $n^2$-qubit register of the form
\[
\cH_q:=(\C^2)^{\otimes n^2}\,,
\]
where qubit $(i,j)$  corresponds to  lattice site $(i,j)$  and 
the basis states $|0\rangle$ and  
$|1\rangle$ 
indicate that there is no atom or that there is an atom, respectively,  
at position $(i,j)$. 
A subspace of $\cH_q$ is 
the clock Hilbert space $\cH_c$ spanned
by all {\it allowed} configurations. They correspond to those basis states
in $\cH_q$ for which there is exactly one atom in each row 
and for which the atom chain is connected. 
First we introduce  interactions between site
$(i,j)$ and $(i+1,j+1)$ 
which 
generates independent hopping of atoms 
along the rows:
\begin{equation}\label{VHam}
K:=\sum^{n-1}_{i,j=1}\Big( a_{i,j}  a^\dagger_{i+1,j+1} + h.c.\Big) \,.
\end{equation}
To achieve that the atom configuration remains a connected chain 
we introduce a strong attractive force between diagonal neighbors: 
\begin{equation}\label{Attract}
H_{pot}:=-E\sum_{<(i,j) , (k,l)>}  N_{i,j} N_{k,l} + E_0\, {\bf 1}\,,
\end{equation}
where $E\in \R^+$ 
is assumed to be much larger than $1$ and $N_{i,j}=a^\dagger_{i,j} a_{i,j}$ is the projection on the states with an atom in position $i,j$. The sum runs 
over all unordered pairs of sites $(i,j),(k,l)$ 
with $|i-k|+|j-l|=1$.
The physically irrelevant term $E_0$ is (for purely 
technical reasons) chosen such
that the initial configuration has zero energy. 
The energy is minimal if all attractive interactions are active. Otherwise,
if the chain is not connected, the potential energy is at least
$E$. 
We define
the ``synchronization Hamiltonian'', which leads
to the desired coupled motion of 
atoms, by 
\begin{equation}\label{SynEq}
\tilde{H}:=K+H_{pot}\,.
\end{equation}
In order to analyze the dynamics generated by $\tilde{H}$ 
 we first argue  
that it can be replaced with an effective clock Hamiltonian
$H_{eff}$ having  $\cH_c$ as an invariant subspace. The initial atom configuration defines a ground state of $H_{pot}$. 
Due to $E\gg 1$   the comparably 
small perturbation $K$ cannot circumvent the gap to one of the first
excited states of $H_{pot}$. 
If $P$ denotes the
spectral projection of $H_{pot}$ corresponding to the ground state energy
$0$ 
we will therefore expect a dynamical evolution according to
\begin{equation}\label{effdef}
H_{eff}:=P K P\,.
\end{equation}
Note that not all states in the image of $P$ are allowed clock states
since the image of $P$ contains also states with more than 
one atom in a row. However, these ground states could only be reached from
an allowed state if $K$ contained hopping terms along columns.

The following lemma (proven in the appendix) justifies
more formally that we may analyze the dynamics according to $H_{eff}$
instead of $\tilde{H}$:

\begin{Lemma}\label{Approx}
For $E> n^6$ and $n\geq 10$ 
the norm distance between the effective Hamiltonian
in equation~(\ref{effdef}) and the restriction $\tilde{H}_{E/2}$ 
of $\tilde{H}$ to the eigenspace corresponding to
eigenvalues  smaller than $E$ 
satisfies
\[
\|\tilde{H}_{E/2}-H_{eff}\|\leq 9 \frac{n^3}{\sqrt{E}}\,.
\]
\end{Lemma}

If we increase
the attractive part of the interaction proportional to $n^6$ 
if $n$ increases the time evolution induced by $H_{eff}$ is a good
approximation for the true one. If the intention of this article was to
propose a  physical implementation scheme this would
be a rather bad scaling, leading possibly
to hard practical problems. However, from the computer science
point of view, it is essential that the increase of energy 
is only polynomial in $n$ and we will thus obtain an {\it efficient}
model of computation.

We will now give a more explicit form
of  $H_{eff}$. To be more precise, we will describe an operator
whose restriction to $\cH_c$ coincides with $H_{eff}$.  
Let $|c\rangle$ denote some basis state in $\cH_c$, 
i.e., $c$ describes an allowed atom configuration. Then
\begin{equation}\label{HeffZ}
H_{eff}|c\rangle=P K P |c\rangle= P K|c\rangle=
P\,\sum^{n-1}_{1= i,j} \Big(a_{i,j}  a^\dagger_{i+1,j+1} + h.c.\Big)|c\rangle
\end{equation}
is a superposition of all atom configurations with connected chain that
can be reached from $c$ by moving 
one atom forward or backwards in its diagonal row. In other words,
each hopping term 
$a_{i,j}  a^\dagger_{i+1,j+1}$
is only active if the sites $i,j+1$ and $i+1,j$ are occupied. 
We have therefore a conditional hopping 
given by
\[
H_{eff}=\sum^{n-1}_{i,j=1} 
a_{i,j}  a^\dagger_{i+1,j+1} N_{i,j+1}N_{i+1,j} + h.c.\,\,\Big|_{\cH_c}\,,
\]
where the symbol $\cH_c$  at the right indicates the restriction to $\cH_c$. 
Using a slightly more intuitive notation, 
which respects the orientation of Fig.~\ref{DiagonalPositionen}, 
the summands of $H_{eff}$ 
can be
denoted by

\begin{equation}\label{4Qu}
\begin{array}{ccccc}
 &          & N &       &  \\
 & \otimes  &   & \otimes  & \\
a &         &   &          &  a^\dagger\\
 & \otimes  &   & \otimes &    \\
& & N & & 
\end{array}  \,\,\,\,+ \,\,\,\,h.c.\,,
\end{equation}
where this $4$-local operator acts on $4$ black sites enclosing a common
white site. 
 
We are interested in  the dynamical evolution 
of the quantum state defined by the initial atom configuration. 
Since it is a ground state of the dominating Hamiltonian $H$ 
the major part of the state vector 
is contained in that spectral subspace of $\tilde{H}$ 
which corresponds to energy values not greater than $E/2$. 
Using the bound of Lemma~\ref{Approx} we may then estimate 
the error in the dynamical evolution caused by replacing the true
evolution with the evolution according to $H_{eff}$.

\begin{Theorem}\label{Fehler}
Let $|\psi\rangle \in \cH_c$ be an arbitrary allowed 
atom configuration.
Then we have 
\[
\|(e^{-i\tilde{H}t} -e^{-iH_{eff}t}) |\psi\rangle \|\leq 
\epsilon t + 2n\,\sqrt{\frac{2}{E}}\,,
\]
where 
\[
\epsilon:=9 \frac{n^3}{\sqrt{E}}
\]
as in Lemma~\ref{Approx}.   
\end{Theorem}

\vspace{0.5cm}
{\it Proof:}
Due to $H_{pot}|\psi\rangle=0$ 
the average energy of $|\psi\rangle$ satisfies
\[
\langle \psi |\tilde{H}|\psi\rangle 
=\langle \psi |K|\psi\rangle 
\leq \|K\| \leq n^2\,,
\]
where the last inequality is given by counting the number of terms in 
eq.~(\ref{VHam}).
Let $Q$ be the spectral projection of $\tilde{H}$
corresponding to all
eigenvalues  less than $E$.  Then we have
\[
\langle \psi | Q\tilde{H} Q|\psi\rangle +
 \langle \psi |({\bf 1}-Q)\tilde{H}({\bf 1}-Q)|\psi\rangle=
\langle \psi|\tilde{H} |\psi\rangle \leq n^2\,. 
\]
Using
\[
\langle \psi |({\bf 1}-Q)\tilde{H}({\bf 1}-Q)|\psi\rangle
\geq E \|({\bf 1}- Q)|\psi\rangle\|^2\,,
\]
and
\[
\langle \psi |Q\tilde{H}Q|\psi\rangle =\langle \psi |K|\psi\rangle 
\geq -\|K\| \geq -n^2
\]
we conclude 
\begin{equation}\label{Rem}
\|({\bf 1}-Q)|\psi\rangle \|^2 \leq \frac{2n^2}{E}\,.
\end{equation}
We have
\begin{eqnarray}\nonumber
\|(e^{-i\tilde{H}t}-e^{iH_{eff} t})|\psi\rangle \|
&\leq& \|(e^{-i\tilde{H}_{E/2}t}-e^{-iH_{eff}t}) Q|\psi\rangle \|\\
&&
+\|(e^{-i\tilde{H}t}-e^{-iH_{eff}t})({\bf 1}-Q)|\psi\rangle \| \nonumber \\
&\leq & \epsilon t +2 \|({\bf 1}-Q)|\psi\rangle \|\,,
 \label{Arrayeq}
\end{eqnarray}
where we have used
$
\|\exp(iA) -\exp(iB)\|\leq \|A-B\|
$
for two self-adjoint matrices $A,B$.
The statement follows then by replacing the last term in 
eq.~(\ref{Arrayeq}) with the right hand side of ineq.~(\ref{Rem}).
$\Box$
\vspace{0.5cm}

One could use the condition
$t\ll \sqrt{E}/(9 n^3)$ of Theorem~\ref{Fehler} 
 to determine the time scale
for which 
$H_{eff}$ is a good approximation
when $E$ is given. We can also 
assume that the time scale is given and we have to 
determine the interaction strength $E$. After all, we have to ensure that
the approximation is valid for all $t\leq t_0$ where $t_0$ is some
a priori given upper bound on the time required by 
the clock wave to pass
the relevant region on the lattice, i.e., the region 
containing the spin-spin 
interactions which will be described in the next section. 

The effective Hamiltonian
leads to a  quantum walk  on the 
space of atom configurations which will
be diagonalized in section \ref{Sec:Diag}.
In the next section we will
explain how this walk may trigger the implementation
of gates.

\section{Holonomic implementation of logical gates}

\label{Holo}

Now we consider not only the positions of the atoms but also
their inner degree of freedom, e.g. their spin.
We replace the 	qubit at each lattice site with a qutrit
and 
 extend hence the space $\cH_q$ to
\[
\cH:=(\C^3)^{\otimes n^2}\,.
\]
The basis states $|0\rangle,|\downarrow\rangle,|\uparrow\rangle$ 
indicate the absence of the atom or its 
two possible spin states, respectively.
Before we describe how to 
add a spin-spin interaction Hamiltonian to $\tilde{H}$ 
which leads
to the implementation of gates when the atoms are moving along the
rows, we should first mention two obvious approaches to imprint
gates by interactions
and why they are not suitable for our goal.

In \cite{Ergodic} and \cite{Margolus:90} 
the gates are directly
coupled (as additional {\it tensor product} components) 
to the synchronization Hamiltonian.  Certainly we could 
define the hopping term in eq.~(\ref{VHam}) 
between column $j$ and $j+1$ 
in any desired row such that 
the spin is not conserved during the atom propagation but
subjected to  some rotation. 
This would provide us with one-qubit gates. Such a one-qubit gate
would clearly be inverted when the atom moves back to column $j$ again.
The 
conditional state of the considered qubit, given that the atom
is found  on the right of column $j$  would hence be subjected
to the desired transformation, no matter whether the atom
has traveled back and forth several times as
is is typical for a random walk. 

To imprint two-qubit gates into the Hamiltonian is more difficult. 
The obvious method to
 couple 
their implementation with some coupled motion of two adjacent atoms
would require more non-local interactions than only
 two-particle terms. 
The second obvious method would be 
to 
complete the clock Hamiltonian
by an {\it additive} 
spin-spin interaction term  between some adjacent sites 
in the same column. This leads to the following problems.
First the atoms do not stay there for a well-defined time. Some part of
the wave packet moves already and one part stays. Second the atoms travel
back and forth and pass the interaction region several times.
Both
effects  would in general entangle atom position and logical spin states
in an uncontrollable way. 

A solution to the latter problem is holonomic quantum computing \cite{PZ00,ZanardiRasetti}.
The idea of this approach, in the usual setting, 
is that some
time-dependent Hamiltonian $G(t)$ 
is adiabatically changed along a
closed loop such that, on an appropriate degenerate 
subspace, the overall effect
is a unitary which depends only on the loop (described by
representing $G(t)$ in some parameter space) and not on the
speed of the change of $G(t)$. The unitary in the end remains the same even
if the vector $G(t)$ has been moving back and forth on this loop in the 
parameter
space.  

We apply this concept to our model where we achieve the 
time-dependence of the spin-spin interaction by 
the motion of the atoms when they pass spatially inhomogeneous
interactions. 
We imprint a spin-spin interaction (varying slowly along the rows)
such that is describes a closed loop 
after
the atoms have passed a certain region. Then the spin state of the atoms
does not depend on the time required for the passing, neither does
it depend on the number of times the atoms were traveling
back and forth before the whole region has been passed.  
This ensures that the entanglement between the 
position degree of freedom and the spin is lost 
as soon as the region has been passed by the whole atom chain. 
Before we explain how to make use of this idea in detail, 
we rephrase  a
result in section II of Ref.~\cite{Majd},  which is useful for us, 
as a lemma:

\begin{Lemma}{\bf (Holonomic Gates)}\\\label{Adia}
Given a family of Hamiltonians $G(t)$  
on a finite dimensional Hilbert space $\cH$ 
by
\[
G(t)=e^{iXt} G_0 e^{-iXt}\,\,\,\,\,\hbox{ with } t\in [0,T]\,, 
\]
where $X$ is a self-adjoint operator on $\cH$ such that
the family $G(t)$ is a closed loop, i.e.,
$G(0)=G(T)$.
Let 
the change of $G(t)$ be sufficiently slowly to consider
it as an adiabatic change, i.e., for each entry $G_{i,j}(t)$ of
$G(t)$ we have 
\[
\Big|\frac{d}{dt} G_{i,j} (t)/G_{i,j}(t)\Big|  \ll \Delta\,,
\]
where $\Delta$ is the smallest spectral gap of $G(0)$.   

Let $U_T$ be the time evolution generated by the family $G(t)$ after 
the closed
loop. Let $\cK$ be an eigenspace of $G_0$ with eigenvalue $\lambda$. 
Then  $U_T$ keeps $\cK$ invariant and its restriction
is given by 
\begin{equation}\label{GeoPh}
e^{-i\lambda T} e^{i QXQ\, T}\,,
\end{equation}
where $Q$ is the projection onto $\cK$.  
The unitary (\ref{GeoPh}) is also
called a non-abelian geometric phase.
\end{Lemma}

Since the holonomic concept implements
the well
defined unitary 
\[
\exp(iQXQ\,T)\] 
only on a {\it subspace} $\cK$ 
of the full Hilbert space 
our $2n-1$ spin particles cannot be used for $2n-1$ logical qubits.
Instead, the atom in row  $2i-1$ will encode a logical qubit
together with the atom in row $2i$. Moreover, 
we can only use rows being not
too far from the row in the middle since the others
are too short to imprint interactions that implement gates. 
Two  adjacent spin particles encode one qubit with  logical states
$|0\rangle_l,|1\rangle_l$ via 
\[
|0\rangle_l:=|\downarrow \rangle \otimes |\uparrow\rangle \hbox{\,\,\, and \,\,\,}|1\rangle_l:=|\uparrow \rangle \otimes |\downarrow\rangle \,.
\]  
We will call their span $\cC$ the ``code space''.

\subsection*{One-qubit gates}

\begin{figure}
\centerline{
\epsfbox[0 0 233 60]{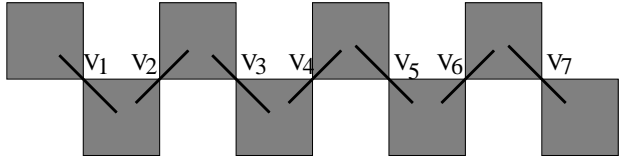}}
\caption{{\small Spin-spin interactions for a logical one-qubit gate. 
Note that these are nearest neighbor interactions in a 2D lattice when the grid is oriented in diagonal direction with respect to the pattern of the 
chess-board.\label{Fig2erWechs}}}
\end{figure}

To explain how to implement
one-qubit gates  
we first restrict our attention  to two adjacent rows.
We add interactions as follows. 
The whole region which carries the gate consists of
a stripe of width $2$ as shown in Fig.~\ref{Fig2erWechs}.
We  call this region ``interaction stripe''.
Inside this stripe,
we add a ``gate Hamiltonian'' 
to  $\tilde{H}$ which consists of spin-spin 
interactions $V_j$  
between certain diagonal neighbors
in adjacent rows (as indicated by the edges labeled with 
$V_1,\dots,V_{l}$ in 
Fig.~\ref{Fig2erWechs}). 
Each $V_j$ is a pair-interaction between two sites. 
We choose the following interactions: 
\begin{equation}\label{Vdef}
V_j:=e^{i\frac{2\pi (j-1)}{l-1} X}
 (\sigma_z \otimes N + N\otimes \sigma_z)   
e^{-i\frac{2\pi (j-1)}{l-1} X}  \,,
\end{equation}
with  $j=1,\dots,l$.
Here $X$ is some self-adjoint operator
on $\C^3\otimes\C^3$ specified later and 
\[
N:=|\downarrow\rangle \langle \downarrow|+
|\uparrow \rangle \langle \uparrow|
\]
is the particle number operator in analogy to the operator $N_{i,j}$ 
in eq.~(\ref{Attract}). 
The Pauli matrix $\sigma_z$ has to be read
as formally acting on the space  $\C^3$ of the corresponding qutrit, even
though it is zero for the state $|0\rangle$ and acts therefore only
on the spin states. 
It is important to note that the degenerate eigenspace of $V_0$
and $V_{l-1}$ coincides with the code space $\cC$. If the spins are initialized 
such that their state
is in $\cC$ before they pass the interaction region 
they will at every time instant remain in the degenerate subspace of the 
interaction that is active (provided that 
the change between different $V_j$ is adiabatic). 
This will implement 
the ``non-abelian phase'' $\exp(i\,QXQ\,T)$ of Lemma~\ref{Adia}.

The length $l$ of the 
interaction stripe is chosen such that 
the change of interactions from $j$ to $j+1$  
can be considered as approximating a continuous change
and furthermore 
as 
adiabatic  when the
atom chain is propagating in horizontal direction.
To sketch how to choose $l$ 
we define a typical time scale $\tau$
on which an atom jumps from one site to its neighbor.
Then we have to ensure that each entry $V_j^{ik}$ of
$V_j$ satisfies
\begin{equation}\label{AdiaAb}
\Big|\frac{V^{ik}_{j+1} - V^{ik}_j}{ \tau \,V_j^{ik}}\Big|  \ll
\Delta\,,
\end{equation}
where $\Delta$ is the smallest energy gap of $V_1$.
Here, the time $\tau$ is defined as a dimensionless
quantity of the order $1$ since the hopping terms
$a_{i,j} a_{i+1,j+1}^\dagger +h.c.$ in  eq.~(\ref{VHam})
appears with constant one.  Physical dimensions 
are irrelevant since inequality (\ref{AdiaAb})
is invariant with respect to  a simultaneous 
rescaling of all $V_j$ and the hopping terms in eq.~(\ref{VHam}). 
We obtain therefore  the dimensionless condition $l\gg 1$.
 
Now we may set
\begin{equation}\label{Xdef1}
X:=\sigma_x \otimes (\cos \phi \,\sigma_x + \sin \phi \,\sigma_z)\,,
\end{equation}
with arbitrary angle $\phi$, 
or
\begin{equation}\label{Xdef2}
X:=\sigma_y \otimes (\cos \phi \,\sigma_x + \sin \phi \,\sigma_z)\,,
\end{equation}
where we have again slightly abused notation since we did not explicitly
indicate 
the embedding of the 2-qubit operators in 
eqs.~(\ref{Xdef1}) and (\ref{Xdef2})  (acting on the 4 possible spin
configurations)
into the two qutrit space. 
The following lemma shows which gates are implemented by the
above choices for $X$.

\begin{Lemma}\label{Onegate}
Let two atoms pass a region where they are subjected to
the Hamiltonians $V_1,\dots,V_l$  in eq.~(\ref{Vdef}) with 
$X$ as in eq.~(\ref{Xdef1}) or eq.~(\ref{Xdef2}). Assume furthermore
that $l$ is sufficiently large to 
consider the change of interactions as adiabatic. 
Then 
their encoded qubit is, up to an irrelevant global phase, 
subjected to
the transformation
$
\exp(2\pi \,i \cos \phi\, \sigma_x)
$
or
$
\exp(2\pi \,i \cos \phi\, \sigma_y)\,,
$
respectively.
\end{Lemma}

\vspace{0.5cm}
{\it Proof:}
On the space $\C^2\otimes \C^2$ spanned by the $4$ spin states
the interaction changes according to 
\[
V_j:=e^{i\frac{2\pi (j-1)}{l-1} X} 
(\sigma_z \otimes {\bf 1} + {\bf 1}\otimes \sigma_z)
e^{-i\frac{2\pi (j-1)}{l-1} X} \,,
\]
with $j=1,\dots,l$. 
Due to Lemma~\ref{Adia} the code space $\cC$ is then
subjected to
$\exp(i\, 2\pi A)$ with
$A:=P_\cC X P_\cC $, where $P_\cC$ is the projection onto $\cC$.
For both choices for $X$, the relevant operator $A$ 
consists then only of 
transitions between $|\uparrow \downarrow\rangle$ and
$|\downarrow \uparrow\rangle$, which corresponds to a $\sigma_x$ 
if the transition amplitudes are $1$ (as it is the case 
given by eq.~(\ref{Xdef1}))
and
to $\sigma_y$ if they are $i$ and $-i$ (as in the case of eq.~\ref{Xdef2}).
$\Box$

\vspace{0.5cm}
Lemma~\ref{Onegate} shows that we can generate arbitrary one-qubit 
transformations by concatenating interactions 
with $X$ as in eq.~(\ref{Xdef1}) and (\ref{Xdef2}) using
appropriate angles $\phi$.

\subsection*{Two-qubit gates}

\begin{figure}
\centerline{
\epsfbox[0 0 233 89]{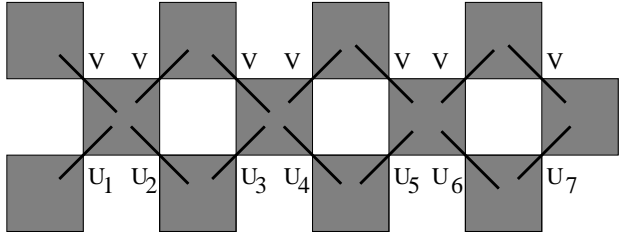}}
\caption{\label{Fig3erWechs}{\small Spin-spin interactions for a logical two-qubit gate. Row 1 and row 2 encode one logical qubit and row 3 is part of
a second qubit (together with row 4 which is not shown in the figure).}}
\end{figure}

For the implementation of 
logical two-qubit gates we consider interaction stripes that consist of
$3$ adjacent rows (see Fig.~\ref{Fig3erWechs}), 
where
row $1$ and $2$  belong to the {\it first} logical qubit and row
$3$ is part of the {\it second} logical qubit. 
We define  interactions
$V$ and $U_j$  such that 
$V$ connects row $1$ and $2$ and $U_j$ connects row $2$ and $3$
as shown in Fig.~\ref{Fig3erWechs}. 
We will refer to the atoms
in row $1,2,3$ as atoms $1,2,3$, respectively.

The interaction between row 1 and the row 2 is 
always $V$ in the whole interaction stripe. Row 2 interacts with  row
3 according to $U_1,U_2\dots,U_l$.

As soon  as atom $2$ has entered the interaction stripe
(shown in Fig.~\ref{Fig3erWechs} for $l=7$)
it is simultaneously subjected to
interactions $V$ and $U_1$. 
Moreover, $V$ and $U_l$ are turned  off simultaneously
as soon as atom $2$ leaves the interaction stripe.
The total interaction that is active on the system
consisting of the spins of atoms 1,2,3 is therefore either 
\[
V \otimes {\bf 1} +{\bf 1} \otimes U_j  \,\,\,\hbox{  with } j=1,\dots,l\,,
\] 
when atom $2$ is inside the interaction stripe, or $0$ otherwise. 

We choose
\begin{equation}\label{V3def}
V:= \sigma_z \otimes N\,,
\end{equation}
where the first tensor component refers to row 1
in Fig.~\ref{Fig3erWechs} and the second to row 2.
The operator $N$ has no effect on the spin of the atom in 
row $2$. It only ensures that the $\sigma_z$-Hamiltonian
is not switched on before the atom in row $2$ has entered 
the left-most black  field in row 3 in  Fig.~\ref{Fig3erWechs}.

The interaction
$U_j$
changes according to 
\begin{equation}\label{Udef}
U_j:= \sigma_z \otimes |\downarrow\rangle \langle \downarrow |
+ (e^{i \frac{2\pi (j-1)}{l-1} \tilde{X} }  \sigma_z e^{-i \frac{2\pi 
(j-1)}{l-1} \tilde{X} })\otimes |\uparrow\rangle \langle \uparrow | \,,
\end{equation}
where $\tilde{X}$ is an operator that acts on the one-qutrit space. It will be
specified later. 
On the space $\C^2\otimes \C^2\otimes \C^2$ 
spanned by all $8$ possible spin states,
we have
\begin{equation*}
V\otimes {\bf 1} +{\bf 1} \otimes 
U_j=\sigma_z \otimes {\bf 1} \otimes {\bf 1} +
{\bf 1}\otimes \sigma_z \otimes  |\downarrow\rangle \langle \downarrow | +
{\bf  1} 
\otimes (e^{i \frac{2\pi (j-1)}{l-1} \tilde{X} }  \sigma_z e^{-i \frac{2\pi 
(j-1)}{l-1} \tilde{X} }       ) \otimes |\uparrow\rangle \langle \uparrow |\,.
\end{equation*}

The idea is that the adiabatic change of the degenerate Hamiltonian
on row $1$ and $2$ is controlled by the spin of the atom in 
row $3$, i.e., by the
logical state of the second qubit. Whenever the state in row $3$ is
$|\uparrow \rangle$, rows $1$ and $2$ are subjected to the 
Hamiltonians 
\[
\sigma_z \otimes {\bf 1} + {\bf 1}\otimes (e^{i \frac{2\pi (j-1)}{l-1} \tilde{X} }  
\sigma_z e^{-i \frac{2\pi (j-1)}{l-1} \tilde{X} })\,,
\]
otherwise they are subjected to a constant Hamiltonian.
We find:

\begin{Lemma}
Let 
$V$ and $U_j$ be as in eq.~(\ref{V3def}) and eq.~(\ref{Udef}), 
respectively, such that they 
connect row $1$ and $2$ as well as $2$ and $3$, respectively.
Choose the corresponding operator 
$\tilde{X}$ as
\[
\tilde{X}:=cos \phi\, \sigma_x + \sin \phi\,\sigma_z\,,
\] 
with arbitrary angle $\phi$. 
Consider the logical 2-qubit space
$\cC\otimes \cC$, where the first
tensor component refers to the atoms
in  rows $1$ and $2$ and the second to those  in 
$3$ and the additional row $4$
(which is not depicted in Fig.~\ref{Fig3erWechs}). 
Then the adiabatic change of the interaction 
$V\otimes {\bf 1} +{\bf 1} \otimes U_j$ 
from $j=1$ to $j=l$ 
 implements a controlled-$\exp(i \sin \phi\, \sigma_z)$ 
transformation on $\cC \otimes \cC$ with the lower qubit as control qubit
and the upper as target (in the tensor product in eq.~(\ref{Udef}) this
correponds to the right and the left side, respectively). 
\end{Lemma}

\vspace{0.5cm}

{\it Proof:}
Let the second qubit be in the logical $0$ state $|0\rangle_l$.
This means that the spins of atom $3$ and $4$ are in the state
$|\downarrow \uparrow\rangle$. However, for the interaction only
the state of atom $3$ matters.
Since it is $|\downarrow\rangle$, atoms $1$ and $2$ are constantly
subjected to the spin Hamiltonian 
\begin{equation}\label{ConstH}
\sigma_z \otimes {\bf 1} + {\bf 1} \otimes \sigma_z\,.
\end{equation}
Since the kernel of this operator coincides with the code space
the logical state is not affected at all. 

Assume now that the second qubit is in the state $|1\rangle$
and atom $3$ is therefore in the state $|\uparrow\rangle$.
Then atom $1$ and $2$ are subjected to the 
Hamiltonians 
\[
\sigma_z \otimes {\bf 1} + {\bf 1} \otimes (e^{i \frac{2\pi 
(j-1)}{l-1} \tilde{X} }  \sigma_z e^{-i \frac{2\pi (j-1)}{l-1} \tilde{X} })\,,
\]
with $j=1,\dots,l$. 
Due to Lemma~\ref{Adia}
the unitary that is 
implemented on the code space after such an adiabatic change 
is given by
$\exp(i2 \pi A)$ with
\[
A:=  P_\cC ({\bf 1}\otimes X) P_\cC= \sin \phi \, P_\cC ({\bf 1}\otimes \sigma_z) P_\cC
\,.
\] 
Describing this operator with respect to the logical basis $|0\rangle_l,
|1\rangle_l$ 
we obtain the operator $\sin \phi \, \sigma_z$. 
This shows that the unitary $\exp(i 2\pi \sin \phi\,\sigma_z )$ 
is implemented after the atoms have passed the interaction stripe. 
Since we have also shown that this unitary is only implemented
if the state of the second qubit is $|1\rangle_l$, we have know shown 
that the net effect is a controlled-$\exp(i 2\pi \sin \phi\,\sigma_z )$
gate.
$\Box$.

\vspace{0.5cm}

We can imprint interactions like those above 
between any adjacent logical qubits 
as well as we may obtain one-qubit rotations as in Lemma~\ref{Onegate}
by appropriate interactions.
It is known that conditional phase gates together with the
set of one-qubit rotations allow for universal quantum computing
\cite{NC}.
Hence we are able to design interactions for  
arbitrary quantum circuits.

\section{The complete Hamiltonian}

\label{Complete}

\begin{figure}
\centerline{
\epsfbox[0 0 240 203]{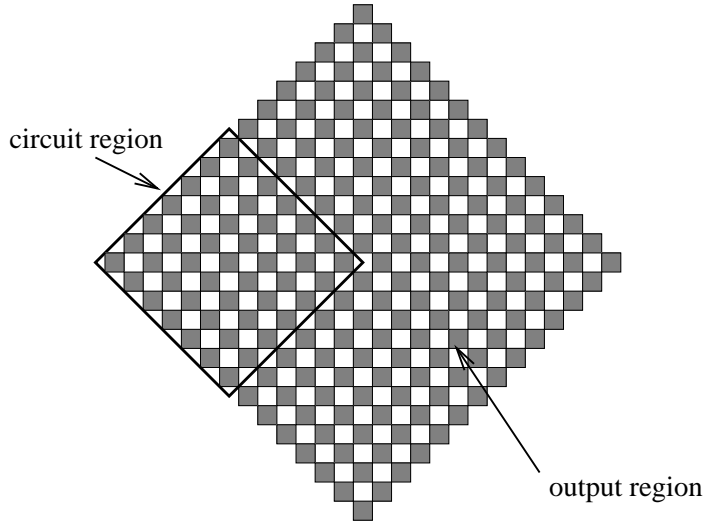}
}
\caption{\label{region} {\small All spin-spin interactions implementing some circuit $U$
are confined to the ``circuit region'' $R$. 
As soon as all atoms have left this
region, the implementation of $U$ is complete.}}
\end{figure}

We obtain the complete Hamiltonian $\tilde{H}$ 
of our autonomous computing device
as a sum of the synchronization Hamiltonian and the spatially inhomogeneous
spin-spin interaction that implement holonomic transformations.
It reads:
\[
\hat{H}:=H_{pot}+K + \sum_{<i,j,i',j'>}W_{i,j,i',j'} \,,
\]
where $H_{pot}$ is the strong attractive force defined in eq.~\ref{Attract}
and
$K$ describes the  hopping terms 
\[
K:=\sum_{i,j=1}^{n-1} a_{i,j,\downarrow} a_{i+1,j+1,\downarrow}^\dagger
+a_{i,j,\uparrow} a_{i+1,j+1,\uparrow}^\dagger +h.c\,,
\]
where we have, in slight abuse of notation, adapted the definition
of eq.~(\ref{VHam}) for spin-less creation
and annihilation operators into those for spin 1/2 particles.
For instance, $a_{i,j,\downarrow}^\dagger$ creates a particle with 
spin down. 
Similarly, the term $H_{pot}$ is adapted to spin 1/2 particles in the sense
that the particle number operator $N_{i,j}$ is given by
\[
N_{i,j}=a_{i,j,\downarrow}^\dagger a_{i,j,\downarrow} +
 a^\dagger_{i,j,\uparrow} a_{i,j,\uparrow}\,.
\]
The spin-spin interactions $W_{i,j,i',j'}$ are only non-zero
if site $(i,j)$ is adjacent to site $(i',j')$ and if the pair of sites 
belongs to some interaction stripe. Inside these stripes, they
are given by the interactions $V,V_j,U_j$ described in the 
preceding section.
We assume that all these rectangles ("interaction stripes") are confined to a
square of length $k$, i.e.,
the spin-spin interaction is only non-zero 
for 
$i,j,i',j' \leq k$ 
for some $k<n$. 
We  call this region  the {\it circuit region}
(see Fig.~\ref{region}).
As soon as all atoms have passed it, the whole circuit is implemented. 
The complement of the interaction region will be called {\it output region}
since we will read out the result of the computation there.

\section{Diagonalization of the quantum walk}

\label{Sec:Diag}
Now we show that the Hamiltonian $\hat{H}$ 
leads to a propagation of atoms that implements the gates in correct order.
First of all, we observe that
the gates are irrelevant for the quantum walk. Different particle
configurations correspond to mutually orthogonal vectors
in the Hilbert space. Whether or not the particles are subjected to an
additional  change
of their inner degree of freedom is irrelevant for the dynamics as long as
we consider the adiabatic limit where the effect of the spin-spin 
interaction is only that it implements {\it unitary gates} 
on the spin states.  Hence we will analyze the  dynamics 
generated by $H_{eff}$ instead of the dynamics generated by $\hat{H}$ 
to derive the relevant probabilities for 
finding the atom chain at certain positions. After recalling  that
the adiabatic change of the spin-spin interactions has implemented
the desired gates  given that the atom chain has passed 
the interaction stripes we have then shown that the circuit has been 
implemented.

To
diagonalize the quantum walk we show that it is isomorphic
to a walk of free fermions propagating on a one-dimensional chain.  
On the left of   Figs.~\ref{DiagonalPositionen} and \ref{Possible<}
we have
characterized the configurations by
binary words of length $2n-2=:2m$. The symbols $0,1$ as  $j$th digit  
indicate whether the atom
in row $j+1$ is {\it behind} the atom on row $j$ or {\it in front} 
of it, respectively.
The initial configuration is hence characterized by $m=n-1$ 
symbols $0$ followed
by $m$ symbols $1$. Since the atom on the very top and that one on the 
very bottom are fixed, the number of symbols $1$ remains constant
and
the vector space  spanned by the possible
atom configurations is therefore the subspace $\cH_m$ of
$(\C^2)^{\otimes 2m}$ spanned by words with Hamming weight $m$. 
With respect to such a representation, the 
effective clock Hamiltonian  $H_{eff}$, when restricted to $\cH_c$, 
 consists of operators which 
replace
some pattern $10$ by $01$ or 
vice versa. It can be written as 
\begin{equation}\label{Hs}
H_s:= \sum_{j=1}^{2m-1} {\bf 1}^{\otimes j-1} \otimes |0\rangle \langle 1|
\otimes |1\rangle \langle 0| \otimes  {\bf 1}^{\otimes 2m-j-1} + h.c.=
\sum_{j=1}^{2m-1} b^\dagger_j b_{j+1} + h.c.\,,
\end{equation}
where $b$ and $b^\dagger$ are fermion annihilation and  creation
operators, respectively. They are
defined \cite{Ar90,ArakiCAR2} by
\[
b_j:=\sigma_z^{\otimes j-1} \otimes |0\rangle \langle 1| \otimes {\bf 1}^{\otimes 2m-j}\,,
\]
and satisfy the canonical anti-commutation relation 
\[
b^\dagger_ib_j+b_j b^\dagger_i=\delta_{ij} \,{\bf 1}\,,
\]
where we have chosen the convention $\sigma_z|1\rangle=-|1\rangle$ and
$\sigma_z|0\rangle=|0\rangle$. 

Observe that 
$H_s$  is the XY-Hamiltonian which is well-known 
in solid states physics \cite{Ar90} and
generates a quasi-free evolution
of fermions. Since the so-called Jordan Wigner transformation,
describing the fermion interpretation formally, is standard 
\cite{ArakiCAR}, we rephrase it only briefly.
The subspace 
$\cH_m$ can be reinterpreted  as the space
of $m$ fermions moving without interactions 
in a $2m$-dimensional state space.
Then $\cH_m$ is simply the antisymmetric tensor product
\[
\cH_m\equiv \underbrace{(\C^{2m} \otimes \cdots \otimes \C^{2m})^-}_{m}\,.
\]
The correspondence between these two pictures can be described easily:
Given a basis state in the qubit register by a binary word
with symbols $1$ at the positions $j_1,\dots,j_m$. 
This state corresponds in the fermionic picture to
\[
(|j_1\rangle \otimes |j_2\rangle\otimes \cdots \otimes |j_m\rangle)^- 
\]
where $|j_i\rangle$ denotes here the $j$th canonical 
basis vector in $\C^{2m}$. 

The restriction of $H_s$ to $\cH_m$ is
in the fermionic picture given by  
\begin{eqnarray}\label{Antisymm}
H_s&=&(S +S^\dagger)\otimes {\bf 1} \otimes \cdots \otimes {\bf 1} \nonumber \\
&+& {\bf 1} \otimes (S+S^\dagger) \otimes {\bf 1} \cdots \otimes {\bf 1}
\nonumber \\
&+&\dots \nonumber \\ 
&+&   {\bf 1}\otimes  \cdots \otimes {\bf 1}\otimes (S+S^\dagger) \nonumber\,,
\end{eqnarray}
where $S$ is the linear (non-unitary) shift 
acting on the basis states of $\C^{2m}$ via 
\[
S|j\rangle := |j+1 \rangle
\]
for $j<2m$ and $S|2m\rangle =0$. This is because the term 
$\sum_j b^\dagger_j b_{j+1}$ shifts the fermion by one site. 
In other words, the time evolution
in each tensor
component in eq.~(\ref{Antisymm}) 
is generated by the Hamiltonian $S+S^\dagger$, i.e., 
the adjacency matrix
of the linear chain. This
implies that the time evolution transfers 
annihilators and creators to linear combinations of annihilators
and creators, respectively, when considered in the Heisenberg picture
\cite{ArakiCAR}.
Let 
\begin{equation}\label{Ut}
U_t:=\exp(-i(S+S^\dagger)t)
\end{equation}
 be the time evolution of one
particle on a quantum walk on a chain and
$u_{jl;t}$ its entries.
Then the time evolution of the creation operator on site $j$ is 
\begin{equation}\label{Heis}
b^\dagger_j \mapsto \sum_{l \leq 2m} u_{jl;t} b^\dagger_l\,,
\end{equation}
i.e., it evolves into an operator creating a fermion
which is in a superposition of different sites.
For the annihilation operator at site $j$ we obtain
\begin{equation}\label{Heis2}
b_j \mapsto \sum_{l \leq 2m} \overline{u}_{jl;t} b_l\,.
\end{equation}
To analyze  the time evolution of relevant observables
we will only make  use of this formulation of the dynamics.

\subsubsection*{Time required for passing the circuit region}

The running time of our computer is given by the length of the
time one has to wait until all atoms will have passed the 
interaction region $R$, i.e., a square of length $k<n$, almost with 
certainty.  
The dynamical evolution $\exp(-iH_s t)$ is clearly quasiperiodic because
the Hilbert space $\cH_m$ is finite dimensional. 
Therefore the atoms will always return to $R$. 
However, the essential idea of the 
ergodic quantum computer 
is that the probability of finding it outside of $R$
is high if one measures at a random time instant. Later, we will therefore
also consider the time average, before we estimate the time 
needed by the atoms to pass $R$ for the first time. 
We found:

\begin{Theorem}\label{PassingTime}
Let the circuit region be a square of length $k$.
Then there is a time $t\in O(k)$  
such that the probability 
of finding all atoms outside the circuit region  $R$ 
is at least  $1-12/k$ given that
the size $n\gg k$ of the whole lattice is sufficiently large. 
\end{Theorem}

The proof can be found in the appendix.
Note that the probability of finding some atom in $R$ could
even be made
smaller if the factor $4$ in eq.~(\ref{jInt}) 
in the proof would be replaced 
with some larger number. 
We conclude that the implementation time for our imprinted 
circuit is in $O(k)$.

\subsubsection*{Computing the time average state}

Now we want to show that also at a random time instant 
the probability of finding all atoms in the output region  (i.e., 
the probability of finding at least $k$ fermions
on the left half) is close to $1$. We found:

\begin{Theorem}\label{PassingProbability}
Let the circuit region be a square of length $k=m/4=(n-1)/4$. 
Then the probability of finding all atoms outside the circuit region 
tends to $1$ for $n\to \infty$.
\end{Theorem}

This theorem is also proven in the appendix.
One should maybe mention the following limitation of our model:
in a lattice with finite length $n$ the adiabatic
approximation underlying the holonomic implementation is only 
true up to some error. Therefore the implemented transformation,
given that the atom chain has passed the circuit region, does not exactly 
coincide with the desired quantum circuit. Instead, it depends to some extent
on the time needed to pass the region. In the limit of averaging 
the time evolution over an infinite time interval this
error will increase. Roughly speaking (and expressed in a too classical
language), the atom chain has then traveled back and forth many times
and the error can increase since passing the circuit region backward
would not necessarily invert the implemented circuit.
However, it is a matter of the
time scale on which the time average is taken whether this error
is relevant, i.e., the degree to which the adiabatic approximation
is justified determines the time scale on which the time average
is described by our analysis.

\section{Remarks on the realization}

\label{Impl}

To judge whether it is likely that systems with Hamiltonians
as above could be found in real systems would go  beyond 
the scope of this article. 
Maybe one should rather ask 
which modifications
are possible for our models that would make them more feasible.
So far, the required spin-spin interactions are  quite specific.

\begin{figure}\label{Dots}
\centerline{
\epsfbox[0 0 221 210]{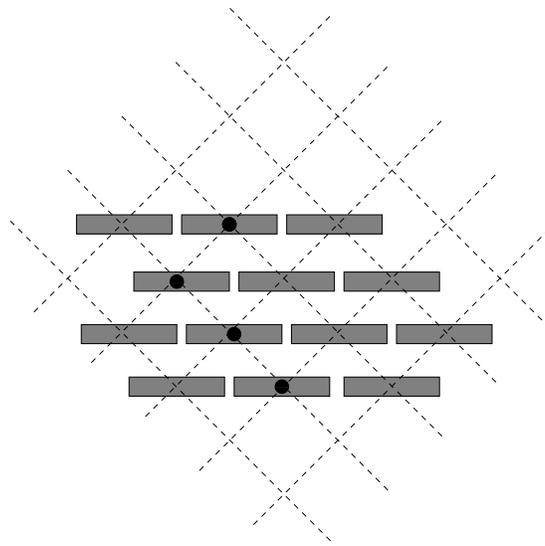}
}
\caption{{\small Schematic drawing of a dot array. 
The black circles indicate the particles carrying the
qubits. Tunneling along
the horizontal lines is possible since the distances are small. 
With respect to the dashed square lattice, the motion is along the diagonal
direction.\label{FigDiaDots}}}
\end{figure}

To show that the required
diagonal hopping is not a priori unphysical
one could think of electrons on quantum dot arrays
arranged as in Fig.~\ref{FigDiaDots}. 
The hopping along the rows does not require tunneling between
distant dots even though it is the {\it diagonal} 
direction with respect to the
square lattice. Spin-spin interactions would then only be needed
in the direction of the dashed square lattice.

Attractive interactions between particles on adjacent dots
could, for instance, be achieved if 
the particles alternate between electrons and holes
from row to row. Certainly, it will be  difficult to find a system
that combines the attractive interaction and the hopping 
with the spatially inhomogeneous spin-spin interactions.

\section{Imprinting a cellular automaton}

\label{Sec:CA}

\begin{figure}
\centerline{\epsfbox[0 0  98 186]{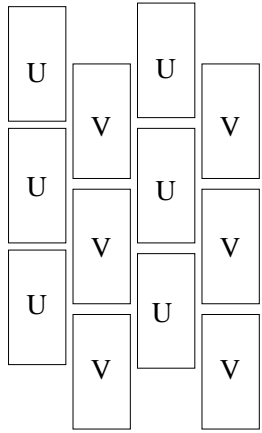}}
\caption{{\small Interaction regions that implement a cellular automaton 
according  to a Margolus partition scheme. A pair of columns simulates
one time step of the CA.}\label{Margo}}
\end{figure}

We have shown that nearest-neighbor interactions among qutrits located
on a 2D square lattice can be designed in such a way that their
autonomous time evolution simulates any desired quantum circuit.
The Hamiltonian given here is significantly simpler than that one given in our
previous paper \cite{Ergodic} since the latter contained $10$-qubit
interactions. However, the advantage of the Hamiltonian 
in \cite{Ergodic} is that
it is 
 programmable. The Hamiltonian given here contains the gates
to be implemented already as hardware. In order
to make a programmable quantum computer
one could proceed as follows.
Consider a universal quantum cellular automaton in one dimension 
\cite{Shepherd} consisting
of cells $C_j$ with $j\in \Z$. Each cell contains a quantum system
with Hilbert space $\C^d$. 
Using the so-called Margolus partition scheme
\cite{WernerCA,Shepherd,RaussendorfCA,Margolus:90}, 
one time step of the CA consists of two 
parts. The first one
is given by
identical copies
of a unitary $U$ acting on all pairs 
$(C_{2j-1},C_{2j})$, the second one by  
identical copies\footnote{To avoid technical problems with infinite
tensor products of the form 
$\cdots \otimes V \otimes V \otimes \cdots $ one could
use a $C^*$-algebraic description and
work with automorphisms on so-called quasilocal algebras
\cite{WernerCA}.} of $V$ acting on 
all pairs $(C_{2j},C_{2j+1})$.
In our scheme, we may represent
each cell by the subspace of an appropriate number of adjacent logical qubits.
Then we 
cover some part of our 2D lattice with a pattern of 
interaction regions that implement
unitaries $U$ and $V$ as shown in Fig.~\ref{Margo}. 
The number of pairs of columns in this pattern determines the number of 
time steps of the simulated CA. 
The
translation symmetry 
of such a crystal would then, however, be described by huge
unit cells.  Nevertheless, our construction shows that
translation invariant finite-range interactions among finite
dimensional quantum systems can in principle 
implement a
universally programmable quantum computer. 

\section{Conclusions}

We have shown that nearest-neighbor interactions on 2-dimensional lattices
can induce dynamical evolutions that are powerful enough to represent 
a programmable quantum computer. Even though we have not described
a realistic implementation scheme we have given 
interactions which are not a priori unphysical. The main reason why 
it may be difficult to construct 
systems having exactly the Hamiltonians considered
here is that the 
spatial homogeneity of the 
interactions follow a rather sophisticated law.

To find even simpler Hamiltonians
which can perform quantum computing in a closed physical system
is an interesting challenge for further research.

\section*{Acknowledgments}

Part of this work was done during a visit of Hans Briegel's group
at IQOQI, whose hospitality is gratefully acknowledged.
The model was inspired by  discussions with 
Robert Raussendorf about cellular automata and with several people
at IQOQI about optical lattices and quantum dots, especially
Gregor Thalhammer and Alessio Recati. 
Thanks also to Pawel Wocjan for encouraging remarks and helpful
discussions. 

This work was partly funded by the Landesstiftung 
Baden-W\"{u}rttemberg, project AZ1.1422.01.

\begin{appendix}

\section{Alternative model avoiding diagonal interactions}

The clocking scheme of our autonomous computer presented so far 
is somewhat sophisticated in the sense that the hopping terms 
connect 
lattice sites 
lying diagonal 
with respect to the grid given by the possible atom
positions. In other words, the rows defining the direction
of atom propagation were {\it diagonal} with respect to the square lattice
structure. It would be more natural to have
tunneling along rows or columns
of the square lattice itself. 

We will therefore describe 
an alternative model for the propagation of an atom chain
which has, however, the 
disadvantage, that we can only conjecture that 
the atoms move in forward direction with sufficient speed and will
appear
outside the interaction region 
with reasonable probability. 
We can, however, prove, that the corresponding classical
random walk  would really  implement the computation. 

\begin{figure}
\epsfbox[0 0 132 191]{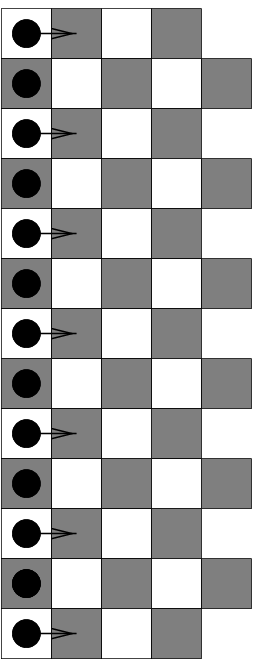} \hspace{1cm} \epsfbox[0 0 205 192]{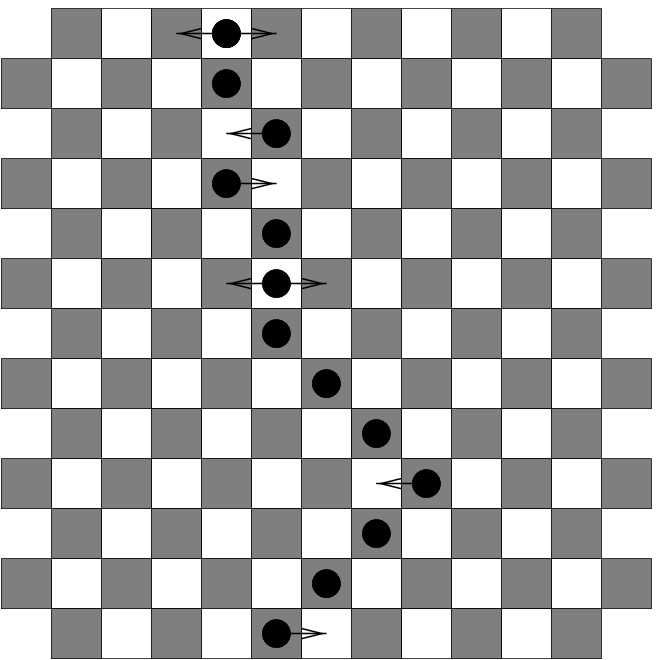}
\caption{\label{Positionen}{\small Left: Possible steps of atoms when they start in the first 
column.
Right: Possible atom configuration and possible motions}}
\end{figure}

The whole lattice consists of $2n\times m$ sites. For reasons that will become
clear later, we will color the sites such that we obtain
a pattern similar to a chess board.
As indicated in
Fig.~\ref{Positionen} the $2n$ atoms start in the first column.
In contrast to our first model, the white sites are used, too.
 
The rules for the  propagation of atoms along the rows 
are as follows. 
An atom on a white site is allowed to move forward
or backwards by one column if the atoms in the two adjacent
rows are currently in the same column. 
An atom on a black site is only allowed to 
move one column forward or backwards if 
it is finally located in between two adjacent atoms in the same column. 
In the first column, only the atoms on the white sites are allowed
to move forward.
A possible configuration obtained by applying these rules is shown
on the left of Fig.~\ref{Positionen}.

By induction, we will argue that the configuration of atoms
has always the following properties:

\begin{enumerate}

\item  Atoms located in adjacent rows are either in the same column or in adjacent columns.


\item For every atom located on a white site there are always atoms on
the two adjacent black sites in the same column.  
The only exception are particles on white sites in 
the upper-most and lower-most rows which have certainly
only one black neighbor. 

\end{enumerate}

It is clear that these conditions are satisfied by the initial state.
In order to prove that  they are preserved we observe that
an atom that is located on a black site 
can only move to a white site 
if the latter is adjacent to  two occupied black sites. 
But then it will satisfy condition (2). 
When an atom on a white site  moves to a black site, condition
(1) is certainly preserved since (2) was true for its initial position.
The atoms being on black sites on the boundary can only move to 
column $j$ when the atom in the row below
is already in column $j$.

Now we present a Hamiltonian that induces these propagation rules.
We start with 
\[
K:=\sum_{k=1}^{2n} \sum_{l=1}^{m-1} a_{k,l}\, a^\dagger_{k,l+1} + h.c.\,,
\]
which is a usual hopping term (as it appears in Hubbard models
\cite{JakschHubbard})
annihilating the atom at a certain 
position and creating it at the right or left neighboring site in the 
same row. 
Physically, one could think
of an optical lattice with 
periodic potentials generated by
standing waves
which result from
the superposition of counter-propagating waves \cite{DuerWalk}. 
To allow tunneling in horizontal direction and to avoid it along 
the vertical
axis one could 
use high potential walls between rows and low potential
walls between columns. 
This can be achieved by 
a superposition of high amplitude 
waves in vertical direction with low amplitude waves  in horizontal direction.

Now we modify the lattice such that there are two types of minima
in the potential. The lower potential corresponds to the black sites
and the higher potential to the white sites. 
By superposition of two lattices with 
wave length $\lambda$ and $2\lambda$
one could also generate  a  chess-board like alternating 
potential (this is called a 
``superlattice'' in \cite{Zoller2Lattice}).

We describe the alternating potential by introducing an additional term
\[
H_1:=-2E \sum_{(k,l)\in B} N_{k,l}\,,
\]
where $B$ denotes the set of black sites. 
Furthermore we introduce an attractive interaction between atoms in adjacent
rows which is only active between pairs of atoms that 
are in the same column. The strength of the attractive interaction 
has strength $E$, i.e., we add a term
\[
H_2:=-E \sum_{k=1}^{2n-1} \sum_{l=1}^m N_{k,l} N_{k+1,l}\,.
\]
This ensures that all the allowed atom configurations
have  the same energy: when an atom sitting on a white 
site moves to a black one the attractive interaction is switched off,
this compensates the energy gap  
between the black and the white site. The same is true
for an atom on a black field that moves to a white field
since the attractive interaction is then switched on. 
For the upper-most and the lower-most rows we need
to decrease the energy gap between black and white 
sites  
because there is only one 
attractive interaction compensating it.
  We achieve this by adding 
a potential 
\[
H_3:=E \,\Big( \sum_{(1,l)\in B}  N_{1,l} +\sum_{(2n,l)\in B} 
N_{2n,l}\Big) \,.
\]
Given the Hamiltonian
$H_1+H_2+H_3$, 
one checks easily that an allowed
forward or backward motion 
 of an atom 
leads to a configuration with the same energy. Moreover,
one can see that every step of an atom that violates the propagation rules
would lead to a state with different energy. 
We observe furthermore that forbidden motions would 
always lead to a configuration with higher energy and never to lower energy.
This additional feature provides the scheme with 
some thermodynamical stability. 

Now we add the small perturbation $K$ ($E$ is again thought to
be large) to the potentials and obtain
\[
\tilde{H}:=H_1+H_2+H_3 +K\,.
\]
In analogy to the discussion of our first model 
we obtain an effective Hamiltonian that is 
given by
\[
H_{eff}:=P \tilde{H} P\,,
\]
where $P$ is projecting onto the subspace of states
having the same energy as the
 initial atom 
configuration. 

The effective Hamiltonian modifies the hopping terms
such that they 
are multiplied by an additional number operator checking 
the position of the atoms in adjacent rows.
In other words, 
each term of the form $a^\dagger_{j,k} a_{j,k+1}$ as well as those
of the form $a_{j,k}a^\dagger_{j,k+1}$ are only active if the
two black fields are occupied which are 
either the current two neighbors of the considered atom
(if the latter is located on a white site) 
or the future two neighbors (if it moves to a white site).
Explicitly, we get therefore
\[
\begin{array}{ccc}
N  & &\\
\otimes & & \\
a  & \otimes &  a^\dagger   \\
\otimes && \\
N & & 
\end{array} 
\hspace{1cm}+\hspace{1cm}
\begin{array}{ccc}
& & N\\
 & & \otimes  \\
a^\dagger  & \otimes &  a   \\
 && \otimes\\
 & & N
\end{array} 
\hspace{1cm}
+
\hspace{1cm}
h.c.\,.
\] 
The operators $N$ of these $4$-local terms
act on  black sites in the same column.

\begin{figure}
\centerline{
\epsfbox[0 0 233 60]{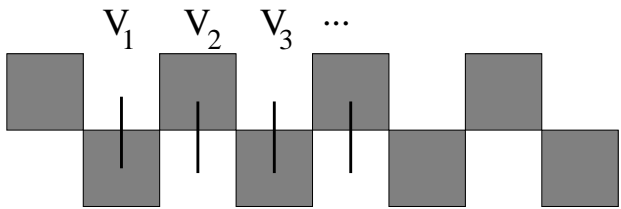}}
\caption{\label{2erFig}{\small Spin-spin interactions for a logical one-qubit gate.}}
\end{figure}

To implement one-qubit gates in this scheme, we imprint 
interactions as shown in Fig.~\ref{2erFig}.
The interactions $V_1,\dots,V_l$ are chosen as in Section~\ref{Holo}.
A decisive difference to our first model is that the hopping of the atoms
does not turn $V_j$ on immediately after $V_{j-1}$ was turned off. Instead,
the atoms pass always a configuration where all interactions are turned off.
However, this is irrelevant for the holonomic scheme because
the  statement of Lemma~\ref{Adia} remains certainly true if
the evolution is interspersed by time intervals
where $G(t)$ is completely switched off.

\begin{figure}
\centerline{
\epsfbox[0 0 233 89]{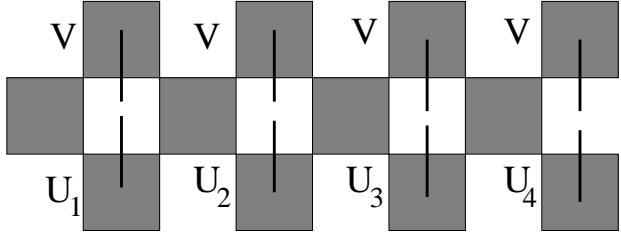}}
\caption{\label{3erFig}{\small Spin-spin interactions for a logical two-qubit gate.
Note that $U_j$ and $V$ are always switched on and  off 
simultaneously since an atom can only visit a white field
if the adjacent black fields in the same column are already occupied.}} 
\end{figure}

To implement two-qubit gates, we design the spin-spin interactions
as depicted in Fig.~\ref{3erFig}.
 Now we have described how interactions can be imprinted that would
implement the desired gates given that the whole atom chain
passes indeed the circuit region.

However,
as already stated, it seems to be difficult to derive explicit 
formulas for the dynamical evolution.
We will therefore only consider the corresponding 
classical random walk and argue that one will have
at least probability
$1/2$ to find all particles outside the circuit region
in the time average. Here the circuit region is defined by 
the first $k$ columns of the chess-board where $k$ is chosen such
that the region contains all interaction stripes.
The 
 computation result is then obtained
by measuring the spin of the particles after they have left the 
circuit region.

The classical random walk would  describe the
physical dynamics in the limit of strongly incoherent atom hopping.
The incoherent model is, of course inconsistent with 
our intention to study computation in a closed physical system.
The motivation to study the classical walk is that the result
gives us
some {\it small hope} that the coherent walk  behaves 
also as it is required for computation even though
it may clearly be very different from the classical random walk. 

Let $C$ be the set of allowed atom configurations on the 
$2n\times m$-chess-board and $G$ be the graph with nodes $C$. 
Two nodes $c_1,c_2 \in  C$ are  adjacent
if $c_2$ can be reached from $c_1$ by an allowed  step of an atom.
The probability distribution on the set of possible atom 
configurations at some
time instant $t$ is described by a vector
 $p(t):=(p_1(t),\dots,p_l(t))$
having the $l$ nodes of $G$ as indices.   
Then a continuous classical random walk on $G$ is described  by \cite{Chung}
\[
\frac{d}{dt} p(t)= Lp (t)\,,
\]
where $L$ is the graph Laplacian. Its   entries are 
$
L_{ij}:= -1 
$
for $i\neq j$ if $(i,j)$ is adjacent, and $L_{j,j}=d_j$ where
$d_j$ is the degree of node $j$, i.e., its number of neighbors. 
Since $G$ is connected, $\lambda_0:=0$ 
is a non-degenerate eigenvalue of $L$ \cite{Chung} 
and there is hence a unique stationary distribution. 
The latter is clearly invariant if we replace each atom configuration
with the configuration obtained by  reflecting 
it 
at the vertical symmetry axis of the chess-board. The latter is defined
by a line between column $2/m$ and 
$1+2/m$ if $m$, where we have assumed that $m$ is even for simplicity. 

Therefore the probability of finding at least one atom of the chain
in the right half is at least $1/2$. Since the chain does not tear
this implies that the column index of every atom is
at least $m/2-2n$. 
Provided that 
$k<m/2-2n$ we find with probability
at least $1/2$ all atoms outside the circuit region.

\section{Proof of Lemma~\ref{Approx}:}

We have a dominating Hamiltonian $H_{pot}$ and a weak perturbation $K$. 
In order to use a perturbation theorem of \cite{Oliveira} 
we first have to rephrase some
notations  used in \cite{Oliveira,Kempe2local}.  

Given a Hamiltonian $H_{pot}$ and a small perturbation $K$, both acting
on some finite dimensional Hilbert space $\cH$. Let $H_{pot}$ have  a
spectral gap $\Delta$ such that no eigenvalues lie 
between $\lambda_-=\lambda_*-\Delta/2$ and $\lambda_+=\lambda_*+\Delta /2$
for some $\lambda_*\in \R$. 
Let $\cH_-$ be the eigenspace of $H_{pot}$ corresponding to
the eigenvalues below $\lambda_*$ and  denote its complement by 
$\cH_+$. Denote the corresponding spectral projections by 
$P_\pm$. For an arbitrary operator $X$, write
$X_{\pm \mp}$ for $P_\pm X P_\mp$ and $X_+$ for $P_+ X P_+$. 
We define the self-energy operator $\Sigma_-(z)$ for real-valued $z$ by
\[
\Sigma_-(z):= H_{pot\,-}+K_{--} + K_{-+} G_+( {\bf 1 }- K_{++} G_+)^{-1}K_{+-}\,,
\]
where $G_+$, called the unperturbed Greens function (resolvent) in the 
physics literature, is defined by 
\[
G_+^{-1}(z):=z \,{\bf 1}_+ -H_{pot \,+}\,.
\]
Using the above notations,
we rephrase the following  
theorem which can be found as Theorem~8  in  \cite{Oliveira}.

\begin{Theorem}\label{ResTh}
Set
$\tilde{H}=H_{pot} +K$ 
with $\|K\|\leq \Delta/2$. 
Let there be an effective Hamiltonian
$H_{eff}$ with spectral width $w_{eff}$. We assume that $H_{eff}=P_- 
H_{eff}P_-$. Let $0<\epsilon<\Delta$ and furthermore

\begin{enumerate} 

\item there is an $r\in \R$ such that
\[
w_{eff} +2\epsilon \ll r \ll \lambda_*\,,
\]

\item for all $z\in \C$ such that $|z|\leq r$, $\|\Sigma_-(z) -H_{eff}\|\leq \epsilon$.

\end{enumerate}

Then the restriction 
$\tilde{H}_{<\lambda_*}$ of $\tilde{H}$ to the eigenspaces with eigenvalues smaller than $\lambda_*$  satisfies 
\begin{eqnarray}\label{Entsch}
\|\tilde{H}_{<\lambda_*}-H_{eff} \|&\leq & 3(w_{eff}+\epsilon)\sqrt{\frac{\|K\|}{\lambda_*+\Delta/2-w_{eff}-\epsilon}}\\
&+&\frac{r^2 \epsilon}{(r-w_{eff}-\epsilon)(r-w_{eff}-2\epsilon)}\nonumber \,.
\end{eqnarray}
\end{Theorem}

We set $\lambda_*:=E/2$ and $\Delta:=E$. Then the condition
$\|K\|\leq \Delta /2$ is
true since the expression in eq.~(\ref{VHam}) is a sum of less
than $n^2$ terms of norm $1$ and we have hence $\|K\|< n^2< E$  due to
the assumption $E\geq n^6$.
For the same reason, the spectral width $w_{eff}$ of $H_{eff}$ 
is less than $n^2$. The operator $P_-$ projects onto $\cH_c$
and condition $H_{eff}=P_-H_{eff}P_-$  required by the theorem 
is satisfied by definition 
(see eq.~(\ref{effdef})).
We will now derive a bound on 
\[
\|\Sigma_-(z)-H_{eff}\|
\]
in order to define an appropriate $\epsilon$ and a corresponding constant $r$.

Due to $K_{--}=P_- K P_-=H_{eff}$ and $H_{pot\,-}=0$ we have
\[
\Sigma_-(z)-H_{eff} =K_{-+}G_+({\bf 1}-K_{++}G_+)^{-1} K_{+-}
\]
The eigenvalues of $H_{pot\,+}$ 
are $E,2E,3E,\dots$ according to the number 
of inactive interactions. Therefore the norm of
\[
G_+(z)=(z\,{\bf 1}_+ + H_{pot \,+})^{-1}
\]
is
at most  $2/E$
 for all $z$ with 
$|z|\leq E/2$. With
\[
({\bf 1} - K_{++}G_+)^{-1}= \sum_{n\geq 0} (K_{++}G_+)^n\,,
\]
we have
\[
\|({\bf 1} - K_{++}G_+)^{-1}\|\leq \sum_{n \geq 0} 
\Big(\frac{2\|K\|}{E}\Big)^n =\frac{1}{1-2\|K\|/E}\,.
\]
Hence we get
\[
\|\Sigma_-(z)-H_{eff}\|\leq \|K\|^2 \|G_+\| \frac{1}{1-2\|K\|/E}\,.
\]
For $E\geq 4\|K\|$ we obtain with $\|G_+\|\leq 2/E$ 
\[
\|\Sigma_-(z)-H_{eff}\|\leq \|K\|^2 \frac{4}{E} \leq 4\frac{n^4}{E} 
=:\epsilon\,.
\]
We may choose $r:=\sqrt{E}$ in order to fulfill
$w_{eff} +2 \epsilon \ll r \ll \lambda_*$. 
We may now use  
inequality~(\ref{Entsch}) and obtain 
\begin{eqnarray*}
\|\tilde{H}_{E/2}-H_{eff}\| &\leq& 3(n^2+1)
\sqrt{\frac{n^2}{E-n^2-1}}\\&+&\frac{4 n^4}{(\sqrt{E}-n^2-1)(\sqrt{E}-n^2-2)}\,,
\end{eqnarray*}
where we have inserted the following results and definitions from
above: $w_{eff}<n^2$, $\lambda_*=E/2$, $\Delta =E$, $r=\sqrt{E}$.
We have inserted $\epsilon=4n^4/E$ only in the numerator of the 
second fraction and replaced it with $1$  (as an upper bound) 
at the other places. 
Using  $E-n^2-1> E/2$ for $E>n^6$ and 
$n\geq 10$ and $\sqrt{E}- n^2 -2 > \sqrt{E}/2$ 
we have
\[ 
\|\tilde{H}_{E/2}-H_{eff}\|\leq 4n^2 \sqrt{\frac{2 n^2}{E}} +
\frac{16 n^4}{E} \leq 9\frac{n^3}{\sqrt{E}}\,, 
\]
where we have used $16 n^4/E <n^6/E < n^3/\sqrt{E}$. 
Thus, we have obtained the desired upper bound on 
the norm distance between $\tilde{H}_{E/2}$ and $H_{eff}$.

\section{Proof of Theorem~\ref{PassingTime}}

Let 
\begin{equation}\label{Idef}
|I\rangle :=|0\dots 01\dots 1\rangle \in \cH_m
\end{equation} 
be the initial configuration when denoted in the qubit picture.
Recall the beginning of section~\ref{Sec:Diag}, where
we have described the correspondence between clock states and binary words
 (as indicated by the sequence
of symbols $1$ and $0$ at the left of Figure~\ref{DiagonalPositionen}).
 
Then all atoms are outside of $R$ if and only if at least $k$ symbols $1$ 
have traveled from the right half of the chain to the left half, i.e.,
at least $k$ fermions are contained in the left half of the interval
of length $2m$. We have therefore to solve a mixing problem of a 
``discrete free fermion  gas'' 
where all particles start in the right interval. 
First we show that after the time $O(k)$ it is likely that
at least $k$ symbols $1$ can be found on the left side.
We define the observable 
\[
\cN:=\sum_{j=1}^m  P_j\,,
\]
where $P_j:=b^\dagger_j b_j$ 
is the projector on the upper state of qubit $j$. 
The observable $\cN$ counts the number of symbols $1$ on the left side.
In Figure~\ref{Possible<} this corresponds to
the number of symbols $1$ above the row in the middle.
We will estimate 
the probability that less than $k$ symbols $1$ have moved to the
left by the Chebyshev inequality.  It states that for any random
variable $X$ we have
\begin{equation}\label{Tsche}
P( |X-E(X)| \geq \epsilon )\leq \frac{V (X)}{\epsilon^2}\,,
\end{equation}
where $E(X)$ and $V(X)$ denote the expectation value and the 
variance of $X$,
respectively. In the sequel we will consider 
the probability distribution of $\cN$ as the distribution of
such a classical random variable.

Let $|I_t\rangle$ be the time evolved state after time $t$. 
Then the expectation value of $\cN$ after the time $t$ is given by
\begin{equation}\label{exp}
E_t(\cN):=\sum_{j=1}^m \langle I_t|P_j|I_t\rangle= 
\sum_{j=1}^m\langle I_t|
b^\dagger_j b_j |I_t\rangle \,.
\end{equation}
Using the dynamics (\ref{Heis}) and (\ref{Heis2}) we get
\begin{eqnarray}\label{u_ij}
\langle I_t| P_j |I_t\rangle &=&
\langle I_t | b^\dagger_j b_j |I_t\rangle 
=\sum_{l=1}^{2m} u_{jl;t}\, \overline{u}_{jl;t} \langle I | b^\dagger_l b_l |I\rangle\\&=&
\sum_{l=m+1}^{2m} u_{jl;t}\, \overline{u}_{jl;t} 
= \nonumber
\sum_{l=m+1}^{2m} |u_{jl;t}|^2\,,
\end{eqnarray}
where we have used that $\langle I|b_l^\dagger b_l|I\rangle$ is $1$ 
or $0$ depending on whether there the state $|I\rangle$ contains
the symbol $1$ at this position. 
Each term $|u_{jl;t}|^2$ is the probability for a particle starting at site
$l$ to be found at site $j$ when measured after the time $t$
in a single particle quantum walk on a linear chain of length $2m$. 
Since the time evolution of such a walk has been discussed in detail in 
the literature \cite{ChildsWalk}, we will only describe the implications
for our model.

Consider a particle starting at site  $l$ with 
\begin{equation}\label{jInt}
m+1 \leq l \leq m+ 4 k\,.
\end{equation} 
In the notation of eq.~(\ref{Idef}) such a fermion corresponds to one
of the $4k$ leftmost symbols $1$.  
The assumption 
$n\gg 4 k$ and hence $m\gg 4k$ ensures that 
the time interval considered in the sequel is sufficiently small 
compared to the time to reach the boundaries of the lattice.
Then 
reflections at the boundaries can be neglected. 
We have to wait only 
the time $O(k)$ in order to achieve that the width of the 
wave function of a particle starting at a definite position
is much larger than $4 k$ (see \cite{ChildsWalk}) but still smaller
than the size $n$ of the whole computer.
Then the probability of finding it on the left half is larger than
$1/3$. 
Recalling that this holds true for
each $l$ satisfying eq.~(\ref{jInt}) and 
that the number of expected fermions
 ($\equiv$ symbols ``$1$'') 
in the left half 
 is given by
the sum in eq.~(\ref{exp}), we may choose
a time instant $t\in O(k)$ such 
that the expectation value satisfies exactly
\[
E_t(\cN)=\frac{4}{3}\,\,k\,.
\]
In order to estimate the variance 
\begin{equation}\label{VarDef}
V_t(\cN)=E_t(\cN^2)-(E_t(\cN))^2\,, 
\end{equation}
we observe that
eq.~(\ref{exp}) implies
\begin{equation}\label{E2t}
(E_t(\cN))^2=\sum_{i,j\leq m} 
\langle I_t| P_i|I_t\rangle \langle I_t| P_j |I_t \rangle\,. 
\end{equation}
Moreover, we have
\begin{equation}\label{Anni}
E_t(\cN^2)=\sum_{i,j\leq m} \langle I_t| P_i P_j |I_t\rangle=
\sum_{i,j\leq m, i\neq j} 
\langle I_t| P_i P_j |I_t\rangle + E_t(\cN)\,,
\end{equation}
where we have used 
\[
\sum_{i=1}^m \langle I_t | P_i P_i |I_t \rangle =\sum_{i=1}^m 
\langle I_t | P_i|I_t \rangle=E_t(\cN)\,,
\]
due to eq.~(\ref{exp}).
We rewrite one summand of the first term on the right of eq.~(\ref{Anni})
 as
\begin{equation}\label{Anni2}
\langle I_t| P_i P_j |I_t\rangle=
\sum_{l,s,r,p=1}^{2m} u_{il;t} \,\overline{u}_{is;t} \,u_{jr;t} \,\overline{u}_{jp;t} 
\langle I|b^\dagger_l b_s b^\dagger_r b_p |I\rangle\,.
\end{equation}
The inner product can only be nonzero if annihilators meet creators, i.e.,
if either $l=s$ and $r=p$ or $l=p$ and $s=r$ or if all indices
coincide. In the first case (including the third) 
the term is only non-vanishing for $l=s>m$ and $r=p>m$ since
$b^\dagger_l b_l$ is the projection $|1\rangle \langle 1|$ on qubit $l$.
In the second case we must have $l=p>m$ and $s=r\leq m$ since
$b_s b_s^\dagger$ is the projection $|0\rangle \langle 0|$ on qubit $s$. 
Hence eq.~(\ref{Anni2}) becomes
\begin{equation}\label{2Term}
\sum_{m<l,\,s\leq 2m} u_{il;t} \,\overline{u}_{il;t} \, u_{js;t}\, \overline{u}_{js;t}
+ \sum_{m< l\leq 2m,\, s\leq m} 
u_{il;t}\, \overline{u}_{jl;t} \,u_{js;t} \,\overline{u}_{is;t}\,.
\end{equation}
The first term coincides with 
$
\langle I_t| P_i|I_t\rangle \langle I_t| P_j |I_t \rangle 
$
by eq.~(\ref{u_ij}).
We rewrite the second term as  
\begin{eqnarray*}
&&\sum_{m\leq l \leq 2m,s \leq m} u_{il;t}\, \overline{u}_{jl;t} \,u_{js;t}\, 
\overline{u}_{is;t}  \\ &=&
 \sum_{1\leq l \leq 2m,\,\, s\leq m} u_{il;t} \,\overline{u}_{jl;t}\, 
u_{js;t}\, \overline{u}_{is}-
\sum_{1\leq l \leq m} u_{il;t}\, \overline{u}_{jl;t}\, \sum_{1\leq s\leq m}
\, u_{js;t} \,\overline{u}_{is;t} 
\\&=&
- |\sum_{1\leq l \leq m} u_{il;t} \, \overline{u}_{jl;t}|^2  
\,,
\end{eqnarray*}
where the last equality is due to 
\[
\sum_{1\leq l\leq 2m} u_{il;t} \overline{u}_{jl;t} =\delta_{ij}\,,
\]
because $U_t$ is unitary (see eq.~\ref{Ut}).
Hence we have found
\begin{equation}\label{kleiner}
\langle I_t| P_i P_j |I_t\rangle\leq 
\langle I_t| P_i|I_t\rangle \langle I_t| P_j |I_t \rangle\,\,\,\,\,\,\forall i\neq j\,.
\end{equation}

Combining (\ref{VarDef}) and (\ref{kleiner}) with (\ref{E2t})
straightforward computation shows  
\begin{equation}\label{Vung}
V_t(\cN)\leq \Big(E_t(\cN)-\sum_{i\leq m} 
(\langle I_t| P_i |I_t\rangle)^2\Big)\leq E_t(\cN)\,.
\end{equation}

Recall
that we have chosen the time instant such that 
$E_t(\cN)=4k/3$ and hence $V_t(\cN)\leq 4k/3$ by ineq.~(\ref{Vung}).
Assume we would find
less than $k$ symbols $1$ on the left half.
Then the random variable defined by 
$\cN$-measurements would deviate at least $k/3$ from its
expectation value. Hence we can apply eq.~(\ref{Tsche}) with
$V_t(\cN)\leq 4k/3$  and  
$\epsilon^2=  k^2/9$ and find that the probability of finding less 
than $k$ symbols in the left half is at most $12/k$.
This completes the proof.

\section{Proof of Theorem~\ref{PassingProbability}}

In analogy to the proof of Theorem~\ref{PassingTime}
we will compute the expectation value and the variance of $\cN$ in the time average
state and then use the Chebyshev inequality (\ref{Tsche}).

The time average expectation value
of the fermion number on the left half  is given by averaging
eq.~(\ref{exp}) over all $t$:
\begin{equation}\label{TAV}
E(\cN):=\lim_{T\to \infty} 
\frac{1}{T}\int_0^T E_t(\cN) \, dt= 
\sum_{l=m+1}^{2m} 
\Big(\sum_{j=1}^{m}\lim_{T\to \infty} \frac{1}{T}\int_0^T |u_{jl;t}|^2 \,dt\Big)\,,
\end{equation}
where
the last equality follows from eq.~(\ref{u_ij}).
Recall that we interpret each summand for $l=m+1,\dots,2m$  
as the probability of finding a particle (that has started at site $l$)
in the left half of the chain when measured at a random
time instant. We will show that it is close to $1/2$ up to 
an error in $O(1/\sqrt{m})$. 
It is known \cite{CDS} that the  Hamiltonian 
$S+S^\dagger$ on $\C^{2m}$ 
generating the walk 
(i.e., the
adjacency matrix of the ``path graph $P_{2m}$'') 
has $2m$ different eigenvalues
\[
\lambda_r = 2 \cos \frac{r \pi}{2m+1}\,\,\,\,\,r=1,\dots,2m\,.
\]
The eigenspaces of $S+S^\dagger$, 
are therefore one-dimensional
and the
 $r$th eigenvector 
is
 given by \cite{Pothen} 
\[
|e_r\rangle:= \sqrt{c} \sum_{j=1}^{2m}
\cos \Big((j-\frac{1}{2})(r-1) \frac{\pi}{2m}\Big) |j\rangle\,,
\] 
where the normalization factor  is $c=1/(2m)$ for $r=1$ and $c=1/m$ 
for $r\neq 1$. 

Using these eigenvectors,
we may write the time average density matrix of a particle that has started
at position $l$ as 
\[
\sum_{r=1}^{2m} |e_r\rangle \langle e_r|l\rangle \langle l|e_r\rangle \langle e_r|\,.
\] 
By
evaluating the probability to be at position $j$ using this state
we may compute the time average 
in eq.~(\ref{TAV}) and obtain
\begin{equation}\label{limqr}
\lim_{T\to\infty} \frac{1}{T}\int_0^T |u_{jl;t}|^2 \,dt =
\sum_{r=1}^{2m}| \langle j|e_r\rangle|^2\, |\langle e_r|l\rangle |^2\,.
\end{equation}
Each eigenvector 
defines a probability distribution $p_r$ on $\{1,\dots,2m\}$ by
\begin{equation}\label{pdef}
p_r(j):=| \langle j|e_r\rangle|^2 =\frac{c}{2} \Big[ 1+ \cos \Big((j-\frac{1}{2})(r-1) \frac{\pi}{m}\Big)\Big]\,.
\end{equation}
The spatial probability oscillations are given by  waves with frequencies
$\nu_r:=(r-1) \pi/m$. We will refer to those frequencies $\nu$  
that satisfy
\[
2\pi \frac{1}{\sqrt{m}} 
\leq \nu \leq  2\pi (1-\frac{1}{\sqrt{m}})
\]
as {\it high} frequencies  (they are neither close to $0$ nor
close to $2\pi$)
and to the others
as {\it low} frequencies. 
For a high frequency wave, 
the probability distribution $j\mapsto p_r(j)$ 
leads almost to equal probabilities
for both halfs.
This is seen from
\[
\sum_{j=1}^m p_r(j)=\frac{1}{2} +\frac{1}{m}\sum_{j=1}^m 
\cos (j \nu_r -\frac{\nu_r}{2})\,.
\]
The sum over the oscillating terms 
can be bounded from above by
observing
\[
|\sum_{1\leq l \leq m} \cos (l\nu -\frac{\nu}{2})|
\leq |\sum_{1\leq l \leq m} e^{il\nu}| 
\leq \frac{2}{|1-e^{i\nu}|} \in O(\sqrt{m})\,,
\]
where we have used that $\nu$ 
differs from both $0$ and $2\pi$ by $\Omega(1/\sqrt{m})$
and
 the last inequality follows directly from the geometric sum
formula 
\[
\sum_{j=0}^n q^j =\frac{1+q^{n+1}}{1-q}\,.
\]
Using eq.~(\ref{limqr}) and eq.~(\ref{pdef}), the term in the bracket in eq.~(\ref{TAV}) can be written as
\begin{equation}\label{Sumj} 
\sum_{j=1}^{m}  \sum_{r=1}^{2m}  p_r(j) \, |\langle e_r | l\rangle |^2\,.
\end{equation} 
Neglecting 
the low frequency terms $p_r$ in this sum can only cause an
error in $O(1/\sqrt{m})$ due to  $|\langle e_r | j\rangle |^2 \in O(1/m)$ 
taking into account that there are $O(\sqrt{m})$ such terms. 
Hence the probability of finding a particle that has started at any site $l$ 
in the left half is close to $1/2$ up to an error in $O(1/\sqrt{m})$.
By summation over all $l=m+1,\dots,2m$ we find $E(\cN)-m/2 \in O(m/\sqrt{m})$. 

To derive bounds on the variance of $\cN$ in the time average state we 
use eq.~(\ref{Anni}) and observe
\begin{eqnarray}\label{firstline}
E(\cN^2)&=& \sum_{1\leq i,j \leq m, i\neq j}
\hbox{average}\Big(\langle I_t | P_i P_j |I_t\rangle\Big) +E(\cN)\\&\leq &
\sum_{1\leq i,j \leq m, i\neq j} 
\hbox{average}\Big(\langle I_t | P_i |I_t\rangle \langle I_t| P_j |I_t\rangle\Big)
  +E(\cN)\,, \label{secondline}
\end{eqnarray}
where the last
 inequality is due to ineq.~(\ref{kleiner}).
We may rewrite the term in the big bracket in equation~(\ref{secondline}) 
using eq.~(\ref{u_ij}) into
\begin{equation}\label{Rein}
\langle I_t|P_i|I_t\rangle \langle I_t |P_j I_t\rangle
= \sum_{l=m+1}^{2m} |u_{il;t}|^2 \sum_{l=m+1}^{2m}|u_{jl;t}|^2\,.
\end{equation}
Recall  that the coefficients 
$u_{il;t}$ are the matrix entries of a unitary that describes
a random walk on a linear chain. 
Note furthermore that the corresponding unitary group $U_t$ is generated
by a real-symmetric Hamiltonian and we have therefore $u_{il;t}=u_{li:t}$.
Thus, we are allowed to interpret $|u_{il:t}|^2$ as the probability for 
finding a particle at position $l$ that has started at position $i$,
which interchanges the original roles of $i$ and $l$. 
This will be convenient in the sequel. 
The product of the sums
on the right hand side of eq.~(\ref{Rein})
can either be interpreted as arising from two independent 
walks of two particles or the walk of one particle in two dimensions.
Explicitly, we consider 
 a
dynamical evolution  in
 $\C^{2m}\otimes \C^{2m}$ 
generated by the Hamiltonian
\begin{equation}\label{2Hamil}
(S+S^\dagger )\otimes {\bf 1} +{\bf 1} \otimes (S+S^\dagger)\,.
\end{equation}
Since the Hamiltonian (\ref{2Hamil}) is the adjacency graph
of the square lattice
\[
\{1,\dots ,2m\} \times
\{1,\dots,2m\}\,,
\] 
we can consider
the right hand of eq.~(\ref{Rein})
as the probability of
finding a particle starting at $|i,j\rangle$ 
in a quantum walk on the square lattice  in the ``target quadrant'' 
$\{m,\dots,2m\} \times \{m, \dots ,2m\}$. 
We want to prove 
that we have sufficient mixing in order to obtain 
$1/4$ 
for the time average of term (\ref{Rein}) 
up to an error in $O(1/\sqrt{m})$.  
We 
proceed similarly as for the  one-dimensional walk 
with the essential difference
that the spectrum of 
the Hamiltonian~(\ref{2Hamil}) 
is degenerate since 
$
|e_r,e_p\rangle:=|e_r\rangle \otimes |e_p\rangle
$
and $|e_p,e_r\rangle$ have the same eigenvalues.
We denote the projection onto their span by
$P_{r,p}$. The rank of $P_{r,p}$ is $2$ for $r\neq p$ and $1$ for $r=p$.
Using these projections, we may compute the time average of
(\ref{Rein}) for fixed $i,j$ by
\begin{equation}\label{2DimTAV} 
\sum_{l,s=m+1}^{2m} \sum_{r,p=1}^{2m} \langle l,s 
| P_{r,p} |i,j\rangle \langle i,j| P_{r,p} |l,s\rangle
= \sum_{l,s=n+1}^{2m} \sum_{r,p=1}^{2m} q_{r,p} (l,s)\, d_{r,p}
\,,
\end{equation}
where we have defined the coefficients
\[
d_{r,p}:= \sum_{l,s=1}^{2m} |\langle l,s| P_{r,p}|i,j\rangle|^2 =
\langle i,j|P_{r,p}|i,j\rangle  
\]
and  probability measures 
 $q_{r,p}$ on the square lattice
by
\[
q_{r,p}(l,s):=\frac{1}{d_{r,p}}|\langle l,s 
| P_{r,p} |i,j\rangle|^2\,. 
\]
Defining the frequencies 
$\nu_r:=(r-1)\pi/n$ 
and $\nu_p:=(p-1)\pi/n$,
the state vector given by normalizing 
 $P_{r,p}|i,j\rangle$ for a given $r,p$ is
a superposition of cosine wave functions with frequency 
vector $(\nu_r,\nu_p)$ with another wave having the vector
$(\nu_p,\nu_r)$. The corresponding probability distributions
$q_{r,p}$  contains then terms with
frequencies 
 $2\nu_r,2\nu_p,
\nu_r-\nu_p, \nu_r+\nu_p$. 
We refer to a term as low frequency term 
whenever at least one of these
frequencies is low. For  each ``high frequency distribution''
$q_{r,p}$ the probability for the target quadrant is $1/4$ up to an 
error in $O(1/\sqrt{m})$. This is seen in straightforward 
analogy to the corresponding argument for the one-dimensional walk.
Moreover, the contribution of the low frequency terms in
(\ref{2DimTAV}) can be bounded from above by the number of
low frequency terms (which is in $O(m\sqrt{m})$)
times the maximal coefficient $d_{l,s}$ of each term
(which is 
in $O(1/m^2)$ due to $|\langle i,j|e_r,e_p\rangle |\in O(1/m)$).
We conclude that the total contribution of all low frequency terms
to (\ref{2DimTAV}) is in $O(1/\sqrt{m})$. Hence 
the total sum in eq.~(\ref{2DimTAV}) is $1/4$ up to an error
in $O(1/\sqrt{m})$. We conclude
\[
\sum_{1\leq i,j \leq m, i\neq j} 
\hbox{average}\Big(\langle I_t | P_i |I_t\rangle \langle I_t| P_j |I_t\rangle\Big) =\frac{m^2}{4} +O(m/\sqrt{m})\,.
\]
Using eq.~(\ref{secondline}) we conclude
\[
E(\cN^2)=m^2/4 + E(\cN)+ O(m/\sqrt{m}) 
\]
and hence
\[
E(\cN^2)=(E(\cN))^2+O(m)\,.
\]
Hence the variance $E(\cN^2)-(E(\cN))^2$ is in $O(m)$ and 
the probability that a measurement of $\cN$ leads to a result
with less than $m/4$ converges to zero 
with $O(1/m)$ 
by the Chebyshev inequality.
Therefore, the probability of finding less than $m/4$ fermions on the
left half 
(i.e., the atom chain has
left the circuit region) 
tends  to zero, too.

\end{appendix}




\end{document}